**Probing the 5f Electrons in Am-I by Hybrid Density Functional Theory**

Raymond Atta-Fynn and Asok K. Ray*

*Physics Department, University of Texas at Arlington, Arlington, Texas 76019*

**Graphical Abstract**

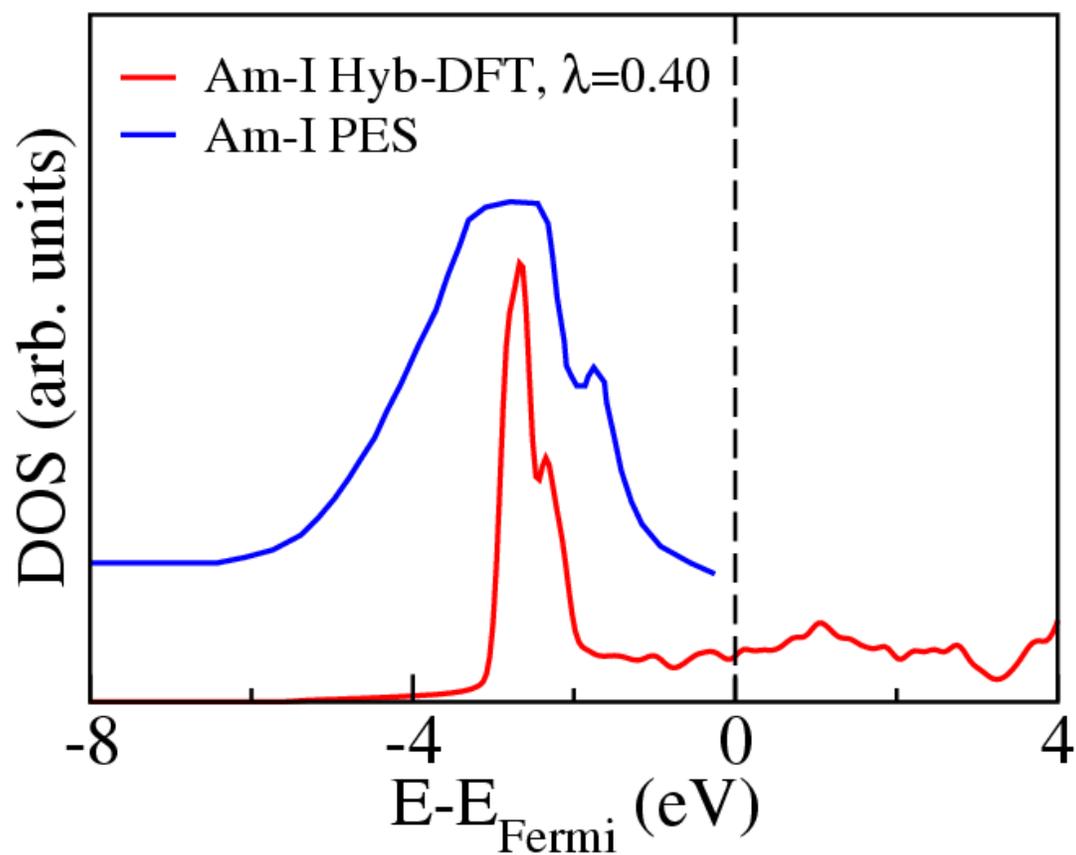

Plot depicting a nice agreement of the computed density of states (DOS) for Am-I using hybrid density functional (40% Hartree-Fock exchange) with photoemission experimental data (blue curve) due to T. Gouder *et al*, Phys. Rev. B **72,** 115122 (2005). The dashed vertical line is the Fermi level.

**Abstract**

The ground states of the actinides and their compounds continue to be matters of considerable controversies. Experimentally, Americium-I (Am-I) is a non-magnetic dhcp metal whereas theoretically an anti-ferromagnetic ground state is predicted. We show that hybrid density functional theory, which admixes a fraction, $\lambda$, of exact Hartree-Fock (HF) exchange with approximate DFT exchange, can correctly reproduce the ground state properties of Am. In particular, for $\lambda$ =0.40, we obtain a non-magnetic ground state with equilibrium atomic volume, bulk modulus, 5$f$ electron population, and the density of electronic states all in good agreement with experimental data. We argue that the exact HF exchange corrects the overestimation of the approximate DFT exchange interaction.

Known for their extraordinary nuclear properties, the electronic structures of the actinides— the group of fourteen elements which follow actinium (Ac) in the periodic table— are characterized by the gradual filling of $5f$ electron states. Actinide elements produce a plethora of interesting physical behaviors due to the $5f$ states. Actinide metal bonding can be separated into two different behaviors, one where the $5f$ electrons are itinerant, that is, they strongly overlap and therefore participate in bonding and one where they offer little or no cohesion and are thus localized. This is evidenced in a parabolic-like decrease in volume with increasing $5f$-electron count (from thorium to plutonium), similar to the early transition metal $5d$ series. This is followed by a large increase in volume from plutonium (Pu) to americium (Am) and little changes for elements beyond, similar to the $4f$ rare-earth series[1-5]. The sudden volume upturn to Am is a signature of $5f$ electron localization. Thus Am, which was first successfully synthesized and isolated at the wartime Metallurgical Laboratory[6], is the first actinide with localized (nonbonding) $5f$ electrons, and as elucidated above, itinerant to localized behavior of the actinide $5f$ electrons occurs somewhere between the different phases of Pu to the phases of Am. Using diamond anvil cell experiments with pressures up to 100 GPa, Heathman *et al.*[7] and Lindbaum *et al.*[8] showed that the crystal structures of Am exhibit the following phase transitions with increasing pressure: double hexagonal close packed (Am I) → face-centered cubic (Am II) → face-centered orthorhombic (Am III) → primitive orthorhombic (Am IV). The last two phases are low symmetry structures with relatively low volume and might be indicative of the participation of the *5f* electrons in bonding.[7,8] Experiments showed that the magnetic susceptibility of Am metal is almost temperature independent, with a $5f^6$ configuration and zero values of spin (*S*), orbital (*L*), and total (*J*) magnetic moments[9-11]. Valence band ultraviolet photoemission data supports the localized character of the Am *5f* electrons in an $f^6$ configuration[12-14].

The most common method for theoretical studies of the electronic and magnetic properties of materials is based on density functional theory (DFT) [15]. Since the local density approximation (LDA) and generalized gradient approximation (GGA) to DFT is based on the homogeneous gas approximation, it works reasonably well in weakly correlated systems. In narrow band systems with localized *d* and *f* electrons, such as the transition metal, rare-earth and actinide compounds, where strong electron correlations are predominant, DFT does not perform very well. In fact, the application of DFT to Am-I and Am II results in magnetic ground states in contradiction with experimental non-magnetic states.[16-20] This failure is due in part to the so called "self-interaction error" (SIE), a spurious term arising from the mean-field Coulomb interaction of an electron density with itself, a consequence of the partial cancellation of the Hartree self-repulsion energy by the self-exchange energy. Thus an accurate description of strongly correlated systems might possibly require methods beyond local and semi-local DFT. One of the methods used to treat strongly correlated materials is the DFT+*U* method, in which an orbital-dependent on-site Coulomb repulsion (via the Hubbard *U*) is added to the DFT Hamiltonian to specifically treat the localized electrons.[21,22] DFT+$U$ calculations with the around-the-mean-field approach for the double counting correction for Am-II yields a non-magnetic ground state but the agreement with experimental photoemission spectra was not very good.[23,24] Another method for treating strongly correlated systems, a merger of DFT and dynamical mean-field theory (DFT+DMFT)[25,26] was used to obtain the correct magnetic and electronic structure of Am[27]. However, $U$ is effectively treated as an *adjustable* parameter in the DFT+$U$ and DFT+DFMT methods and hence they are not strictly *ab initio* formalisms.

Another approach which goes beyond local DFT is *hybrid* density functional theory (HYB-DFT) [28], in which the exchange functional is represented as an admixture of exact non-local Hartree-Fock (HF) exchange with approximate local DFT exchange but the DFT correlation functional is retained. In HF theory, there is no SIE because the Hartree self-repulsion energy is exactly cancelled by the Fock exchange interaction. Hence, SIE in HYB-DFT functional is reduced due to the addition of HF exchange and subsequently, significant improvements in electronic structure properties, *e.g.* band gaps and magnetism of localized $d$ and $f$ electron systems (compared to LDA/GGA) can be expected. HYB-DFT has proved to work reasonably well in *certain* strongly correlated systems for which DFT has failed to perform well. However, the success of HYB-DFT has been mixed at best and *appears* to be system-dependent. For example the PBE0[29,30] hybrid functional (an admixture of 25% HF exchange with PBE exchange-correlation functional[31]) yielded experimental anti-ferromagnetic insulating ground state of $UO_2$ while DFT predicted a ferromagnetic metal[32]. Also the experimentally measured non-zero band gap of $UO_2$ was correctly obtained whereas DFT yielded a zero band gap. Quite recently, HYB-DFT with the PBE exchange-correlation functional was used by us to study the ground state properties of δ-Pu and results, particularly the atomic volume and electronic spectra, showed significant deviations from experimental data at the expense of producing a non-magnetic ground state[33]. The purpose of this communication is to explore the structural, electronic, and magnetic properties of Am-I using HYB-DFT for comparisons with experimental as also standard DFT results.

In the simplest form, a hybrid XC functional $E_{XC}^{HYB}$ containing a fraction $\lambda$ of HF exchange is expressed by:

$$E_{XC}^{HYB} = \lambda E_X^{HF} + (1 - \lambda)E_X^{DFT} + E_C^{DFT} \tag{1}$$

where the subscripts $X$ and $C$ denotes the exchange and correlation terms respectively. Several flavors of hybrid density functional exist in literature. In addition to the PBE0 hybrid functional mentioned previously, other widely used hybrid functionals are the B3LYP[34] and HSE[35]. Here however, since we seek to improve the description of strongly correlated electrons, the hybrid functional is applied but only to a selected set of electrons inside the muffin tin, namely the ones that are poorly treated by DFT, which in this work will be the Am $5f$ electrons. This scheme, which has been implemented in WIEN2k[36-38], leads to significant reductions in computational cost. Specifically, the hybrid exchange-correlation energy functional, $E_{XC}^{HYB}$, used in this work, has the form

$$E_{XC}^{HYB}[\rho] = E_{XC}^{PBE}[\rho] + \lambda \left( E_X^{HF}[\Psi_{5f}] - E_X^{PBE}[\rho_{5f}] \right), \tag{2}$$

where $E_{XC}^{PBE}$ is the Perdew-Burke-Ernzerhof (PBE) formulation of the GGA[31], $E_X^{HF}$ is the HF exchange functional, $E_X^{PBE}$ is the PBE exchange functional, $\Psi_{5f}$ and $\rho_{5f}$ are the wave function and the corresponding electron density of the 5*f* electrons respectively, and $\rho$ is the total electron density. Here, $\lambda$ denotes the fraction of HF exchange replacing the PBE exchange for the Am-I 5*f* electrons. There is no well-defined procedure for determining $\lambda$. $\lambda$ =0.25 is used in most calculations based on perturbation theory arguments[39]. However, we employed several values of $\lambda$ so that we can fully access the evolution of Am-I properties over a wide range of HYB-DFT functional. We should note that using different values of $\lambda$ amounts to the construction of different functional and therefore, $\lambda$ is not an adjustable parameter, making HYB-DFT fully *ab initio*. On the other hand, as mentioned above, DFT+*U* and DFT+DMFT are not *ab initio* because U is adjusted within the theory.

As mentioned above, HYB-DFT total energy calculations were performed using the full-potential augmented plane wave plus local basis (FP-APW+lo) method as implemented in the WIEN2k program. A muffin tin radius of $R_{MT}$=2.9 a.u. was used. The quality of the APW+lo basis set was determined by $R_{MT}K_{MAX}$=9.0. The Brillouin zone was sampled in a mesh of 15×15×4 *k* points. The double hexagonal closed packed structure of Am-I corresponds to four atoms in the unit cell with two atoms at position 2a (0, 0, 0; 0, 0, 1/2) and the other two at position 2d (2/3, 1/3, 1/4; 1/3, 2/3, 3/4) with a space group *P*6$_3$/*mmc* [40]. The non-magnetic (NM), ferromagnetic (FM), and anti-ferromagnetic (AFM) configurations were considered. The anti-ferromagnetic (AFM) configuration was designed by aligning the spins at positions 2a and 2d upwards and downwards respectively.

Scalar relativistic corrections were included the using the approximation developed by Koelling and Harmon[41] to include the mass-velocity and Darwin *s*-shift but omits the spin-orbit terms by performing a *j*-weighted average of the Dirac equations for a spherically symmetric potential field $V(r)$. However, all computations included spin-orbit coupling (SOC), which was added via a second variational scheme: (i) In the first step, the scalar relativistic equations are solved to obtain the eigenstates. (ii) In the second step, the scalar relativistic eigenstates below a certain energy energy cut-off (5 Ry in this work) are used as basis functions for the SOC calculations. Relativistic local orbitals were included in the basis functions to account for the finite character of the $p_{1/2}$ wave function at the nucleus[42]. Also, orbital polarization was included to enhance the orbital magnetic moments following the prescription of Eriksson *et al*.[43] (magnetic DFT calculations usually underestimates the orbital moments in metals).    The spin quantization axis was along the [001] direction. Cell volume optimizations were performed with the ratio of the lattice constants to the experimental value of c/a=3.24.The self-consistent field iterations were terminated when the total energy and charge density converged to $10^{-5}$ Ry and $10^{-3}$, respectively or better.

In Table I, the results of our calculations are summarized. In the fourth column, total energy differences between non-magnetic and magnetic structures are presented for different values of $\lambda$ in the 0.0 to 1.0 range. For the standard DFT approximation ($\lambda$ =0) and HYB-DFT with $\lambda$ up to 0.27, the AFM configuration is more energetically stable, followed closely by the FM configuration, with the NM configuration being the least stable. For values of $\lambda$ in the range 0.29–1.0, the order of stability is reversed, i.e., the NM configuration, the experimentally observed magnetic structure, is preferred. The total magnetic moment (spin plus orbital) per Am atom is presented in the fifth column. The non-zero magnetic moments of the AFM and FM configuration is simply due to the partial cancellation of the spin moment by the orbital moment, with the spin moment being dominant. For the AFM and FM cases, the spin moment is ~6 $\mu_B$/atom for all values of $\lambda$ but the orbital moment gradually increases for increasing $\lambda$ until it saturates for $\lambda \geq 0.2$, leading to a net moment of ~ 3.2 $\mu_B$/atom and ~ 3.5 $\mu_B$/atom, respectively.

The angular momentum coupling scheme in a material (LS, $jj$, or intermediate) is determined by the competition between spin-orbit interaction and the exchange interaction. If the exchange interaction dominates, the coupling scheme is $LS$ (Russell-Saunders), leading to a magnetic solution with maximum spin polarization and splitting of the up- and down-spin bands. In atomic Am in an $f^6$ configuration, this corresponds to $S=3=L$ and thus $J=L-S=0$. If the spin-orbit interaction is dominant, then the coupling scheme is $jj$ which leads to the splitting of the relativistic $j=5/2$ and $j=7/2$ sub-bands. In atomic $f^6$ Am, $jj$ coupling leads to $J=0$ ($L$ and $S$ are no longer good quantum numbers) with the $j=5/2$ sub-shell containing the six $5f$ electrons and the $j=7/2$ shell being empty[24]. If the spin-orbit and exchange interactions have comparable strengths, then the coupling scheme is intermediate. Clearly in the atomic limit, $J=0$ in either the LS coupling or the $jj$ coupling schemes but experimental data shows the presence of no magnetic moments and thus the coupling scheme in Am must be $jj$ or intermediate which is closer to the $jj$ limit. In fact, electron energy-loss spectroscopy (EELS), x-ray absorption spectroscopy (XAS), and spin-orbit sum rule analysis clearly showed that the coupling of the 5$f$ states in Am is intermediate and closer to the $jj$ limit[44,45]. Thus the reason why the magnetic solution is obtained for 0$\leq \lambda \leq$0.27 is due to the overestimation of the exchange interaction, resulting in a large spin polarization and hence large magnetic moments. The inclusion of the exact HF exchange effectively reduces the strength of the exchange interaction, and ultimately has a strength comparable to that of the spin-orbit interaction (and hence the coupling scheme is somewhat intermediate and approaches the $jj$ limit if spin-orbit coupling dominates), leading to a non-magnetic ground state.

In the sixth column of Table I, the 5$f$ electron population inside the muffin tin sphere is reported. In atomic Am with a localized $f^6$ configuration, $n_{5f}$ =6. The fact that $n_{5f}$ > 6 for $\lambda$ =0 implies that $f$ states hybridize with the valence 7$s$ and 6$d$ states, implying that there is some delocalization in the $f$ electron density. With the introduction of HF exchange, the $f$-electron count initially decreases, stabilizing to 5.8. This constant count is clearly reflected in the constant magnetic moment since $f$-electron count is proportional to the spin magnetic moment.

The equilibrium atomic volume and bulk modulus are presented in the last two columns. Shown in parenthesis is the expansion or contraction of the volume relative to experiment. The energy-volume curves for selected values of $\lambda$ are shown in Fig. 1. Clearly the best agreement of the theoretical volume occurs for the AFM configuration ground state, which has already been observed in previous DFT calculations. The NM cell volume shows a large contraction of ~14% initially for $\lambda$ =0 but with the introduction of HF exchange, it slowly expands up to 15% when the full HF exchange is employed. The onset of the volume expansion clearly indicates that the *f* electrons begin to localize at the atomic sites. The gradual expansion of the volume with increasing $\lambda$ can also be observed for the AFM configuration. The trend in volume expansion and contraction for the FM configuration is oscillatory. For $\lambda$ =0.29 and higher, where NM configuration begins to dominate in stability, the volume expands by 4.3%—7.9% for $\lambda$ =0.29—0.40 which is comparable to the 7% error obtained in DFT+DMFT calculations[5]. Surprisingly, the bulk moduli agree with experiment for $\lambda$ =0 but stabilizes to a constant value of 42 GPa for NM magnetic ground state which is of particular interest. It is very clear why the onset of volume expansion leads to an increase in the bulk modulus.

Having demonstrated that the experimentally observed NM ground state exists for $\lambda$ ≥0.29, we now examine the total and the *5f* electron density of states (DOS) and compare them to the experimental photoemission data of Gouder *et al*.[14]. In Figs. 2—4, the total and *5f* DOS for the NM, AFM, and FM configurations are shown for selected values of $\lambda$ , namely $\lambda$ =0, 0.29, and 0.40. Clearly for $\lambda$ =0 case in Fig. 2 for the NM configuration, small splitting exists between the occupied and unoccupied states and a there is fair amount of DOS at the Femi level, implying significant delocalization. Band splitting is observed for the FM and AFM configuration in Fig. 2 but the *5f* states are quite broad, implying delocalization. With a progression from $\lambda$ =0.29 to $\lambda$ =0.40 in Figs. 2–4, it can clearly be seen that that the *j*=5/2 and *j*=7/2 sub-bands of the 5f states of all the configurations are well separated by ~ 4 eV. We verified that the *j*=5/2 band is nearly full and *j*=7/2 band is nearly empty. Also, the center of gravity of total and *5f* DOS of NM is well defined and the localization of the bands increase with increasing $\lambda$ . Most importantly the match of the DOS with experiment is very good. Particularly noteworthy is the fact that the peak shapes, particularly the small one at ~1.8 eV, are well reproduced for $\lambda$ =0.4 in Fig. 4. The total and 5f DOS for the magnetic configurations also localize with increasing $\lambda$ but they are not well-defined as that of the NM configuration. It should be noted that HYB-DFT can be loosely considered as a first order correction to the DFT Hamiltonian in a fraction $\lambda$ of the bare Coulomb interaction and this fraction can be interpreted as a screened Coulomb interaction in a manner similar to the *U* parameter in LDA+*U* calculations[46]. Thus increasing $\lambda$ implies increasing *5f* localization, as can clearly be seen in Figs. 2–4. We verified this for the DOS for $\lambda$ =0.55 and $\lambda$ =1.00 and the states were further localized but failed to match experiment. The volume expansion with increasing $\lambda$ is a direct consequence of the 5f electron localization.

In summary, we have demonstrated, using hybrid density functional theory, that the reduction in the exchange interaction (which is overestimated by approximate DFT) due to the the addition of the exact HF exchange leads to the correct non-magnetic ground state of Am. Specifically, an NM ground state is realized if a fraction of 0.29 or more of HF exchange replaces approximate DFT exchange. For $\lambda$ =0.4, the structural properties are in fair agreement with experiment and also, the computed density of states is in fair agreement with experimental photoemission data due to Gouder *et al.*[14]

This work is supported by the Chemical Sciences, Geosciences and Biosciences Division, Office of Basic Energy Sciences, Office of Science, U. S. Department of Energy (Grant No. DE-FG02-03ER15409) and the Welch Foundation, Houston, Texas (Grant No. Y-1525). This research also used resources of the National Energy Research Scientific Computing Center, Office of Science, U.S. Department of Energy (Contract No. DE-AC02-05CH11231) and the Texas Advanced Computing Center (TACC).

**References**

1. L. R. Morss, N. M. Edelstein, J. Fuger (Eds.) and J. J. Katz (Hon. Ed.), *The Chemistry of the Actinide and Transactinide Elements*, Vol. **1-5** (Springer, New York 2006).

2. D. K. Shuh, B. W. Chung, T. Albrecht-Schmitt, T. Gouder and J. D. Thompson (Editors), *Actinides 2008-Basic Science, Applications, and Technology*, Vol. **1104** (Proceedings of the Materials Research Society, 2008).

3. D. Hoffman (Ed.) *Advances in Plutonium Chemistry 1967-2000* (American NuclearSociety, La Grange, Illinois and University Research Alliance, Amarillo, Texas, 2002).

4. M. J. Fluss, D. E. Hobart, P. G. Allen, J. D. Jarvinen (Eds.) *Proceedings of the Plutonium Futures – The Science 2006 Conference*, J. Alloys. Comp. **444-445** (2007).

5. K. T. Moore and G. van der Laan, *Rev. Mod. Phys.* **81**, 235 (2009).

6. *The Elements beyond Uranium,* G. T. Seaborg and W. D. Loveland, pp. 17, (John Wiley & Sons, Inc. 1990).

7. S. Heathman, R. G. Haire, T. Le Bihan, A. Lindbaum, K. Litfin, Y. Méresse, and H. Libotte, *Phys. Rev. Lett.* **85**, 2961 (2000).

8. A. Lindbaum, S. Heathman, K. Litfin and Y. Méresse, Phys. Rev. B. **63**, 214101 (2001).

9. M. B. Brodsky *Rare Earths and Actinides*, pp. 75 (IOP, London 1971).

10. B. Kanellakopulos, A. Blaise, J. M. Fournier, and W. Müller, Solid State Comm. **17**, 713 (1975).

11. P. G. Huray, S. E. Nave, and R. G. Haire, J. Less-Com. Met. **93**, 293 (1983).

12. J. R. Naegele, L. Manes, J. C. Spirlet, and W. Müller, Phys. Rev. Lett. **52**, 1834 (1984).

13. L. E. Cox, J. W. Ward, and R. G. Haire, Phys. Rev. B **45**, 13239 (1992).

14. T. Gouder, P. M. Oppeneer, F. Huber, F. Wastin, and J. Rebizant, Phys. Rev. B 72**,** 115122 (2005).

15. P. Hohenberg and W. Kohn, Phys. Rev. **136**, B864 (1964).

16. M. Pénicaud, J. Phys. Cond. Matt. **14,** 3575 (2002); *ibid*, **17**, 257 (2005).

17. P. Söderlind and A. Landa, Phys. Rev. **72**, 024109 (2005).

18. D. Gao and A. K. Ray, Eur. Phys. J. B **50**, 497 (2006).

19. O. Eriksson and J. M. Wills, Phys. Rev. B **45**, 3198 (1992).

20. A. L. Kutepov, and S. G. Kutepova, J. Magn. Magn. Mat. **272-276,** e329 (2004).

21. V. I. Anisimov and O. Gunnarsson, Phys. Rev. B **43**, 7570 (1991).

22. V. I. Anisimov, J. Zaanen, and O. K. Andersen, Phys. Rev. B **44**, 943 (1991).

23. A. Shick, L. Havela, J. Kolorenc, V. Drchal, T. Gouder, and P. M. Oppeneer, Phys.

Rev. B **73**, 104415 (2006).

24. V. I. Anisimov, A. O. Shorikov, and J. Kuneš, J. Alloys. Comp. **444-445**, 42 (2007).

25. V.I. Anisimov, A.I. Poteryaev, M.A. Korotin, A.O. Anokhin and G. Kotliar, J. Phys: Condens. Matter **9,** 7539 (1997).

26. A. Georges, G. Kotliar, W. Krauth and M.J. Rozenberg, Rev. Mod. Phys. **68 13** (1996).

27. S. Y. Savrasov, K. Haule, and G. Kotliar, Phys. Rev. Lett. **96**, 036404 (2006).

28. A.D. Becke, *J. Chem. Phys.* **98**, 1372 (1993).

29. M. Ernzerhof and G. E. Scuseria, *J. Chem. Phys.* **110**, 5029 (1999).

30. C. Adamo and V. Barone, *J. Chem. Phys.* **110**, 6158 (1999).

31. J. P. Perdew, K. Burke, and M. Ernzerhof, *Phys. Rev. Lett.* **77**, 3865 (1996).

32. K. N. Kudin, G. E. Scuseria, and R. L. Martin, *Phys. Rev. Lett.* **89** 266402 (2002).

33. R. Atta-Fynn and A. K. Ray, Europhys. Lett. **85**, 27008 (2009).

34. A. D. Becke, J. Chem. Phys. **98**, 5648 (1993).

35. J. Heyd, G. E. Scuseria, and M. Ernzerhof, J. Chem. Phys. **118**, 8207 (2003).

36. P. Blaha, K. Schwarz, G. K. H. Madsen, D. Kvasnicka, and J. Luitz, *WIEN2k, An Augmented Plane Wave Plus Local Orbitals Program for Calculating Crystal properties* (Vienna University of Technology, Austria 2001).

37. Novák P., Kuneš, J., Chaput L. and Pickett W. E., *Phys. Status Solidi B*, **243** (2006) 563.

38. Tran F., Blaha P., Schwarz K. and Novák P., *Phys. Rev. B*, **74** (2006) 155108.

39. J. Perdew, M. Ernzerhof, and K. Burke, J. Chem. Phys. **105**, 9982 (1996).

40. R. W. G. Wyckoff, *Crystal Structures V*ol. 1 (Wiley, New York, 1963)

41. D. D. Koelling and B. N. Harmon, J. Phys. C **10**, 3107 (1977).

42. J. Kuneš, P. Novák, R. Schmid, P. Blaha, and K. Schwarz, Phys. Rev. B **64**, 153102 (2001).

43. O. Eriksson, B. Johansson and M. S. S. Brooks, Phys. Rev. B **41**, 9095 (1990).

44. K. T. Moore, G. van der Laan, R. G. Haire, M. A. Wall, A. J. Schwartz, and P. Söderlind, Phys. Rev. Lett. **98**, 236402 (2007).

45. K. T. Moore, G. van der Laan, M. A. Wall, A. J. Schwartz, and R. G. Haire, Phys. Rev. B **76**, 073105 (2007).

46. D. Jacob, K. Haule, and G. Kotliar, Europhys. Lett. **84**, 57009 (2008).

Table 1.  Equilibrium properties of Am-I for different density functionals. $\lambda$ is the fraction of exact HF exchange, $\Delta E$ is the total energy per atom relative to the non-magnetic configuration ($\Delta E = E - E_{NM}$), $M_J$ is the total (spin+orbital) magnetic moment, $n_{5f}$ is the 5$f$-electron population inside the muffin-tin sphere, $V$ is the atomic volume (the number in parenthesis denotes the deviation of $V$ from the experimental atomic volume), and $B$ is the bulk modulus.

| | $\lambda$ | | $\Delta E$ (mRy/atom) | $M_J$ ($\mu_B$/atom) | $n_{5f}$ | $V$ (a.u.$^3$) | $B$ (GPa) |
|---|---|---|---|---|---|---|---|
| | 0 | NM | 0.0 | 0 | 6.1 | 169.5 (-14.4%) | 36 |
| | 0.10 | NM | 0.0 | 0 | 5.9 | 170.0 (-14.1%) | 48 |
| | 0.15 | NM | 0.0 | 0 | 5.8 | 184.2 (-7.0%) | 40 |
| | 0.20 | NM | 0.0 | 0 | 5.8 | 195.7 (-1.2%) | 44 |
| | 0.25 | NM | 0.0 | 0 | 5.8 | 202.4 (2.2%) | 42 |
| | 0.27 | NM | 0.0 | 0 | 5.8 | 204.6 (3.3%) | 41 |
| | 0.29 | NM | 0.0 | 0 | 5.8 | 206.6 (4.3%) | 42 |
| | 0.31 | NM | 0.0 | 0 | 5.8 | 208.3 (5.2%) | 42 |
| | 0.35 | NM | 0.0 | 0 | 5.8 | 210.8 (6.5%) | 41 |
| | 0.40 | NM | 0.0 | 0 | 5.8 | 213.6 (7.9%) | 42 |
| | 0.55 | NM | 0.0 | 0 | 5.8 | 219.3 (10.8%) | 40 |
| | 1.00 | NM | 0.0 | 0 | 5.9 | 227.9 (15.1%) | 46 |
| | | | | | | | |
| | 0 | AFM | -59 | 5.0 | 6.2 | 197.1 (-0.5%) | 31 |
| | 0.10 | AFM | -119 | 4.4 | 6.0 | 201.1 (1.6%) | 32 |
| | 0.15 | AFM | -93 | 3.5 | 5.9 | 200.3 (1.2%) | 38 |
| | 0.20 | AFM | -62 | 3.2 | 5.8 | 204.4 (3.2%) | 40 |
| | 0.25 | AFM | -25 | 3.2 | 5.8 | 207.9 (5.0%) | 40 |
| | 0.27 | AFM | -9 | 3.2 | 5.8 | 208.6 (5.4%) | 40 |
| | 0.29 | AFM | 7 | 3.2 | 5.8 | 209.7 (5.9%) | 41 |
| | 0.31 | AFM | 23 | 3.2 | 5.8 | 210.6 (6.4%) | 41 |
| | 0.35 | AFM | 57 | 3.1 | 5.8 | 212.1 (7.1%) | 40 |
| | 0.40 | AFM | 100 | 3.2 | 5.8 | 213.8 (8.0%) | 40 |
| | 0.55 | AFM | 233 | 3.2 | 5.8 | 217.7 (9.9%) | 41 |
| | 1.00 | AFM | 640 | 3.2 | 5.8 | 223.6 (12.9%) | 43 |
| | | | | | | | |
| | 0 | FM | -55 | 5.9 | 6.2 | 216.6 (9.4%) | 27 |
| | 0.10 | FM | -115 | 5.5 | 6.1 | 217.1 (9.6%) | 27 |
| | 0.15 | FM | -87 | 4.1 | 5.9 | 211.8 (7.0%) | 31 |
| | 0.20 | FM | -56 | 3.6 | 5.8 | 211.7 (6.9%) | 34 |
| | 0.25 | FM | -20 | 3.5 | 5.8 | 213.3 (7.7%) | 36 |
| | 0.27 | FM | -5 | 3.5 | 5.8 | 214.4 (8.3%) | 36 |
| | 0.29 | FM | 11 | 3.5 | 5.8 | 215.4 (8.8%) | 36 |
| | 0.31 | FM | 28 | 3.5 | 5.8 | 215.9 (9.0%) | 36 |
| | 0.35 | FM | 61 | 3.5 | 5.8 | 217.4 (9.8%) | 36 |
| | 0.40 | FM | 104 | 3.5 | 5.8 | 218.7 (10.5%) | 37 |
| | 0.55 | FM | 235 | 3.5 | 5.8 | 221.0 (11.6%) | 38 |
| | 1.00 | FM | 640 | 3.6 | 5.8 | 224.5 (13.4%) | 40 |

| Expt. | NM | | 0 | ~6 | 198 | 29 |
|-------|----|--|---|----|-----|----|

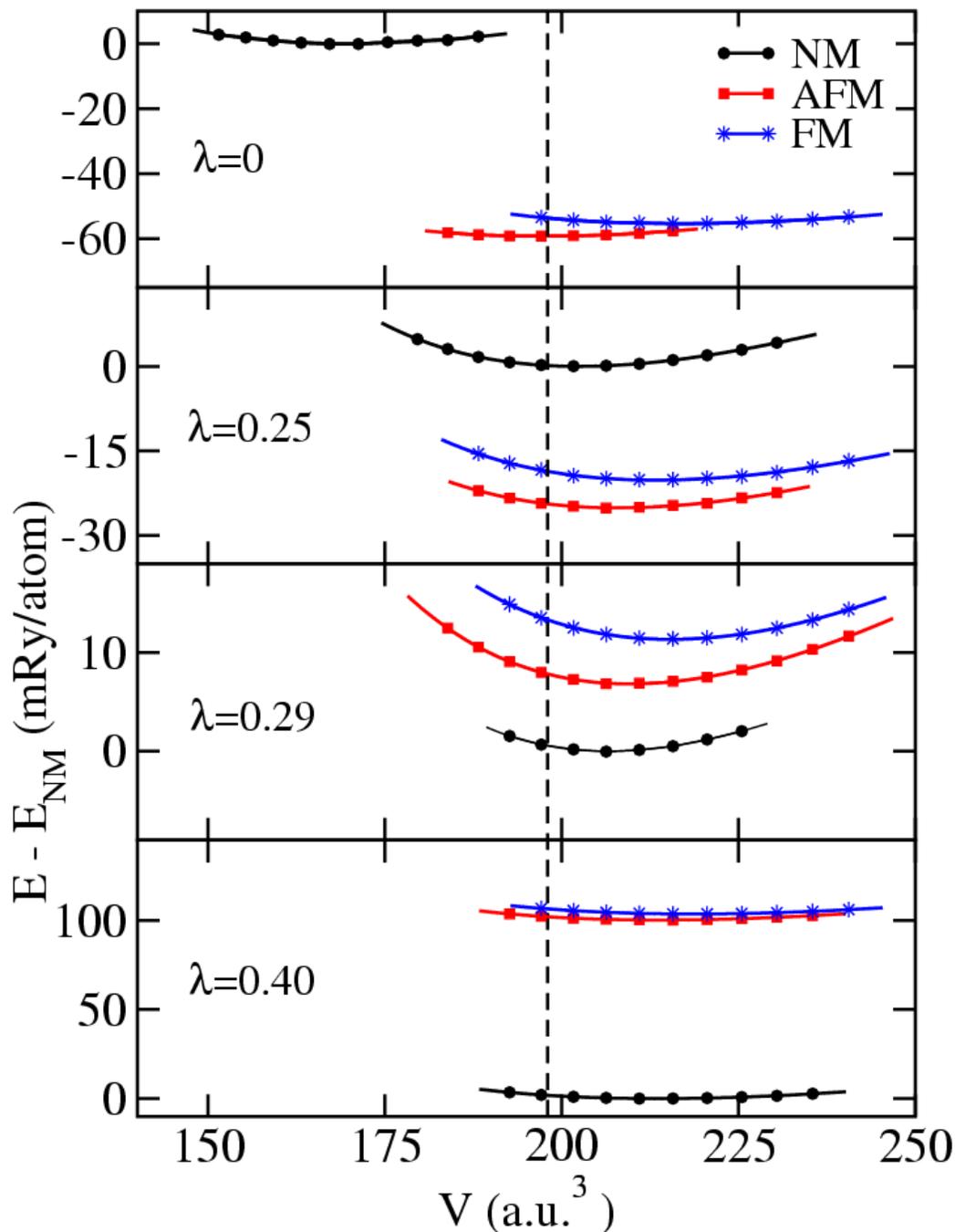

Figure 1: Energy versus volume for $\lambda$ =0.0, 0.25, 0.29, and 0.40 for the non-magnetic (NM), ferromagnetic (FM), and anti-ferromagnetic (AFM) Am-I configurations.

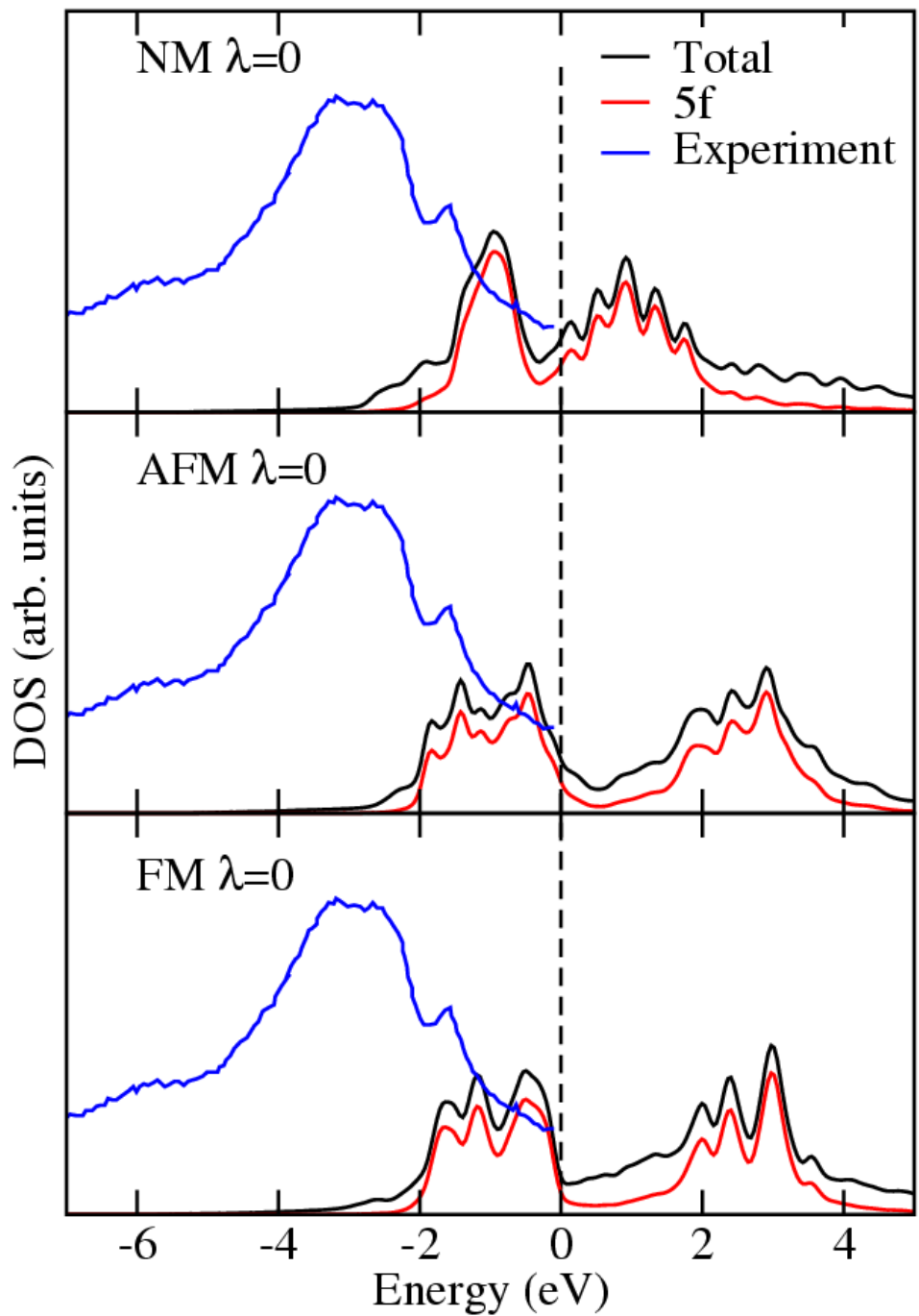

Figure 2: Total and *5f* density of states (DOS), and experimental valence band ultraviolet photoemission data due to Gouder *et al*. (Ref. 14) for Am-I for $\lambda$ =0 (pure DFT functional). The dashed vertical line is the Fermi level.

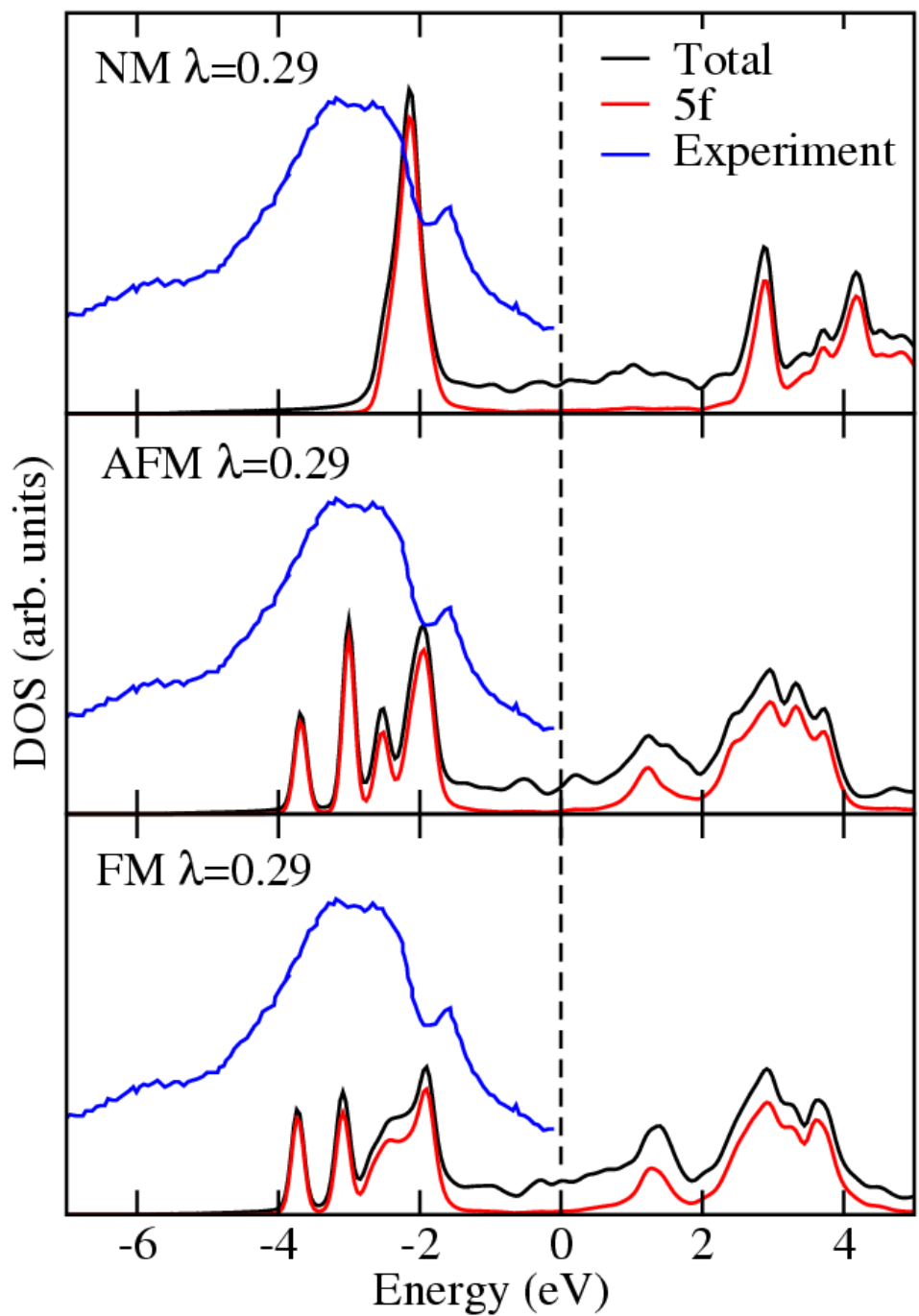

Figure 3: Total and 5f DOS, and experimental valence band ultraviolet photoemission data for Am-I for $\lambda$ =0.29. Note that for this value of $\lambda$ and higher, the NM cell has the lowest energy. The dashed vertical line is the Fermi level.

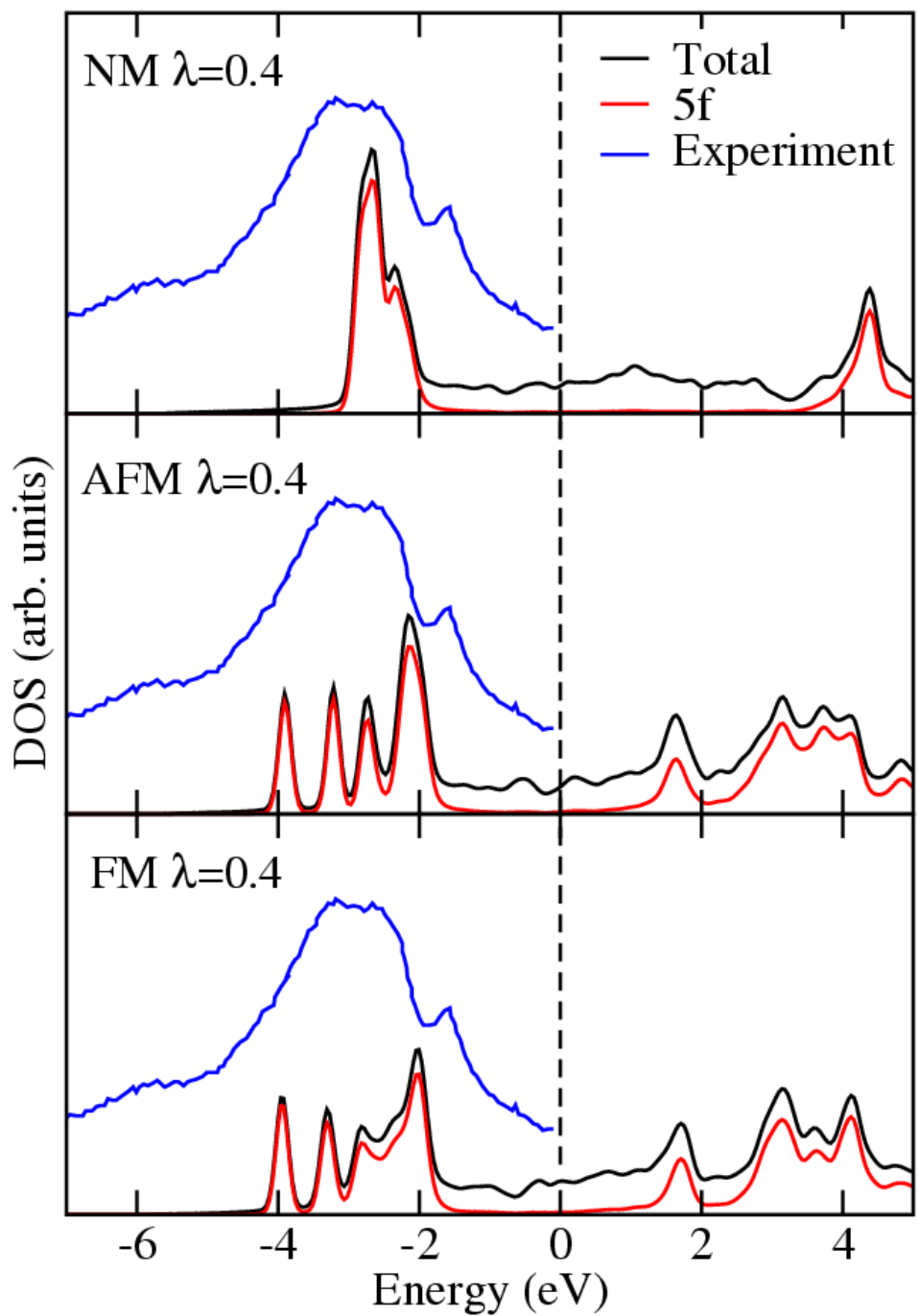

Figure 4: Total and *5f* DOS, and experimental valence band ultraviolet photoemission data for Am-I for $\lambda$ =0.40. The dashed vertical line is the Fermi level.